\documentstyle[epsfig]{mn}

\begin{document}

\title[Pulsations in LS 992/RX J0812.4--3114]
    {Discovery of X-ray pulsations in the Be/X-ray binary \\
LS 992/RX J0812.4--3114}

\author[Reig \& Roche]
{Pablo Reig$^{1,2}$ and Paul Roche$^3$\\
$^{1}$Foundation for Research and Technology-Hellas. 711 10 Heraklion. Crete. 
Greece. \\
$^{2}$Physics Department. University of Crete. 710 03 Heraklion. Crete. Greece\\
$^{3}$Astronomy Centre. CPES. University of Sussex. BN1 9QJ. UK
}

\date{Accepted \\
Received : Version Date of current version \\
In original form ...}

\maketitle

\begin{abstract} 

We report on the discovery of X-ray pulsations from the Be/X-ray
system LS 992/RX J0812.4--3114 during an RXTE observation.  From a
timing analysis of the source we obtained a barycentric pulse period
of 31.8851$\pm$0.0004 s.  The pulse profile is highly structured and
departs from a pure sinusoidal shape.  It shows a sharp dip that may
indicate absorption by the accretion flow.  The energy spectrum from
3-30 keV can be fitted by a power-law model with an exponential
cut-off in accordance with other X-ray pulsars.  The X-ray luminosity
is estimated to be $\sim$ 1.1 $\times$ 10$^{36}$ erg s$^{-1}$ in the
energy range 3-30 keV, assuming a distance of $\sim$9 kpc.

\end{abstract}


\section{Introduction}

The X-ray source RX J0812.4--3114 was discovered by Motch et al.
(1997) during the ROSAT galactic plane survey by cross-correlating the
position of low-latitude X-ray sources ($\mid$b$\mid$ $<$ 20$^{\circ}$) with
SIMBAD OB star catalogues.  The maximum X-ray luminosity in the energy
range 0.1-2.4 keV was reported to be 1.3 $\times$ 10$^{35}$ erg
s$^{-1}$, with some evidence for short-term (hours) variability.
Follow-up optical observations confirmed the identification of source
with the star LS 992.  The optical spectra revealed a B0-1III-V
companion, showing H$\alpha$ in emission (M97). Optical photometry of
LS 992 by Reed (1990) gives V=12.4, B--V=0.41 and U--B=--0.69.  This
gives a reddening free Q parameter of --0.99, implying a spectral
class of B0 or earlier (Halbedel 1993). These data are all consistent
with the source being a new Be/X-ray binary.

Be/X-ray binaries (BeXRBs) consist of a neutron star orbiting a Be
companion.  Accretion onto the compact star is the principal source of
X-ray emission in the system.  A Be star is an early type, luminosity
class III-V star, which at some time has shown emission in the Balmer
series lines.  This emission, as well as the characteristic infrared
excess, is attributed to the presence of circumstellar material, most
likely forming a disc around the equator of the Be star.  The X-ray
emission is characterised by the presence of flares interspersed with
inactivity periods, in which this high energy emission lies below the
threshold of the detectors.  When the flares are modulated with the
orbital period, the increase in flux is typically a factor $\leq$ 10.
Sometimes a giant and unpredictable X-ray outburst takes place with an
increase in flux of 100-1000 above quiescent level.  Thus BeXRBs are
also termed massive X-ray transients (see Negueruela 1998 for a recent
review of BeXRBs).

\subsection{X-ray observations}

LS 992/RX J0812.4--3114 was observed with the {\em Proportional Counter
Array} (PCA) onboard the {\em Rossi X-ray Timing Explorer} (RXTE) on
1998 February 1 and 3.  The total on-source time was 25 ks.
The PCA covers the lower part of the energy range, 2-60 keV, and
consists of five identical coaligned gas-filled proportional units
(PCU), providing a total collecting area of $\sim$ 6500 cm$^2$, an
energy resolution of $<$ 18 \% at 6 keV and a maximum time resolution
of 1$\mu$s.  For a more comprehensive description of the RXTE PCA see
Jahoda et al. (1996).

Good time intervals were defined by removing data taken at low Earth
elevation angle ($<$ 8$^{\circ}$) and during times of high particle
background. An offset of only 0.02$^{\circ}$ between the source
position and the pointing of the satellite was allowed, to ensure that
any possible short stretch of slew data at the beginning and/or end of
the observation was removed. These screening criteria allowed us to
divide the observations up into continuous sections of clean data, on
which the X-ray analyses were carried out.

\section{Timing analysis}

\begin{figure}
    \begin{center}
    \leavevmode
\epsfig{file=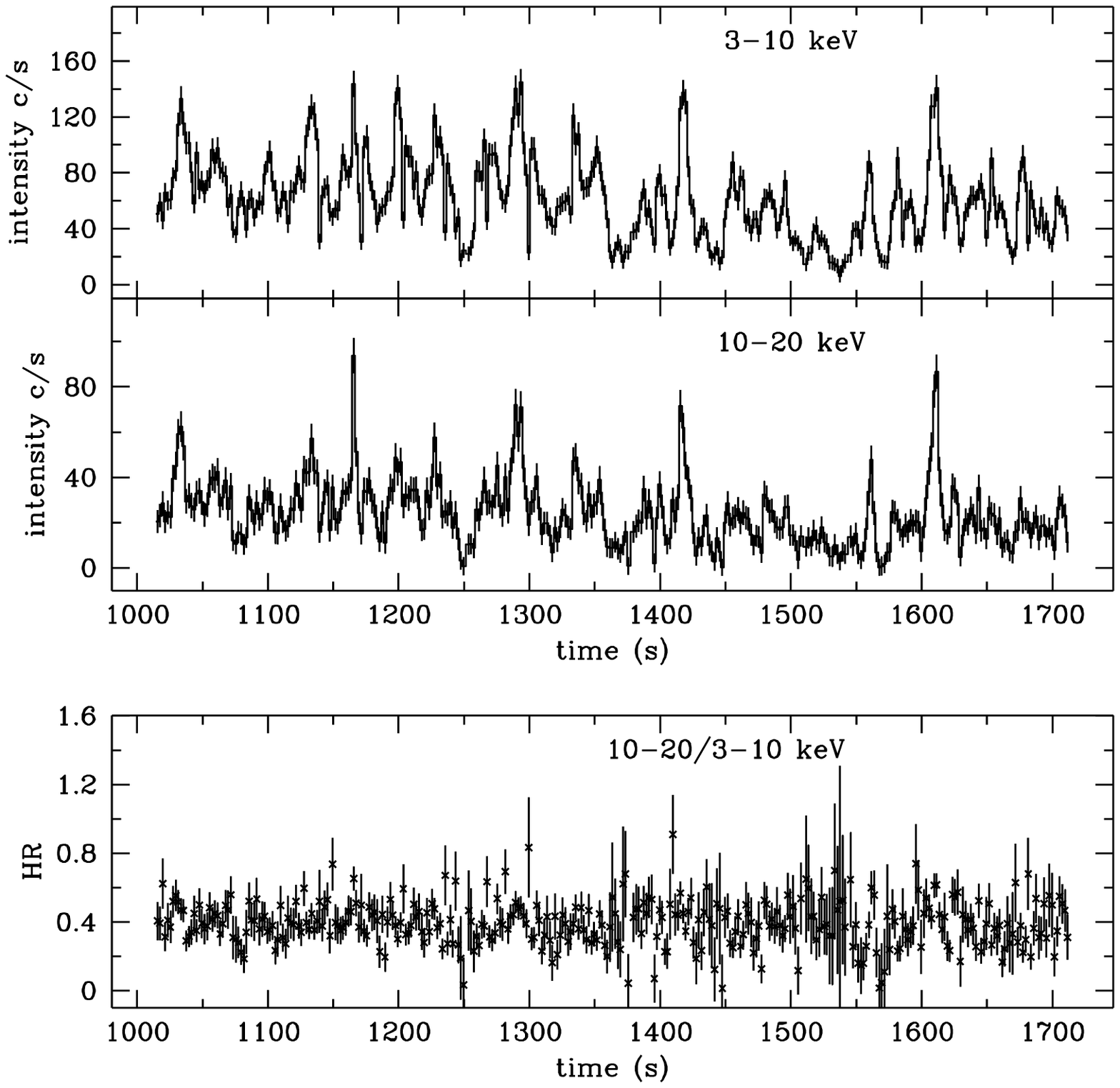, width=8cm, bbllx=54pt, bblly=242pt,
  bburx=545pt, bbury=713pt, clip=}
 \end{center}
        \caption{Lightcurve of LS 992/RX J0812.4--3114 at two
different energies, with each bin representing 2s. Pulsations are
clearly seen at the energies considered. The starting time is JD
2,450,347.045. Also shown is the hardness ratio between 10-20 keV and
3-10 keV.}
        \label{ls992_lc}
        \end{figure}

\begin{figure}
    \begin{center}
    \leavevmode
\epsfig{file=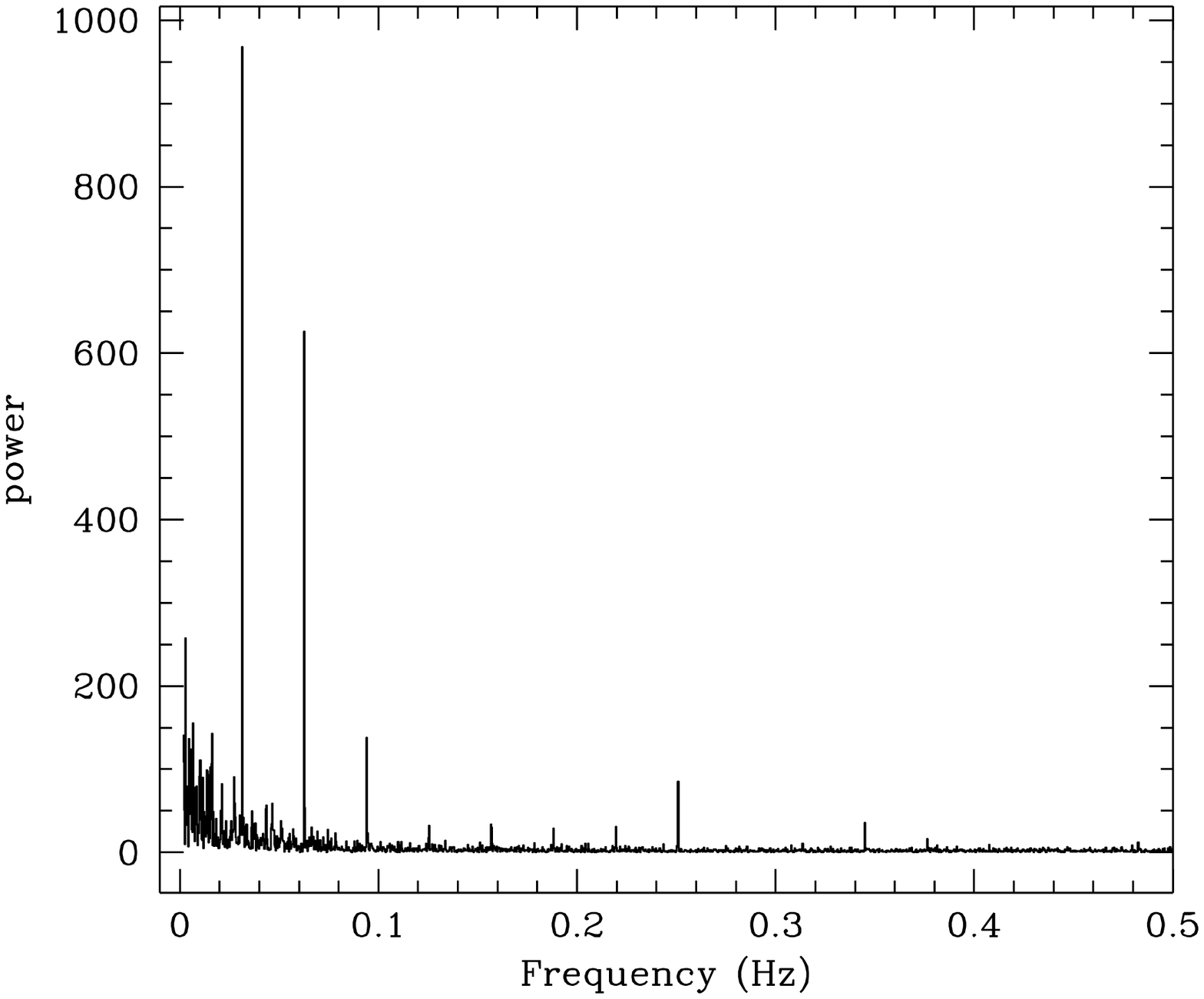, width=8cm, bbllx=20pt, bblly=216pt,
  bburx=530pt, bbury=655pt, clip=}
 \end{center}
        \caption{Power density spectrum of LS 992/RX J0812.4--3114 showing the
presence of coherent modulation with high harmonic content at $\nu
\approx 0.031$ Hz.}
        \label{pds992}
        \end{figure}

The X-ray lightcurves at 3-10 and 10-20 keV of LS 992/RX J0812.4-3114
are shown in Fig \ref{ls992_lc}, together with the time variatiability 
of the hardness ratio 10-20/3-10 keV.  The mean PCA count rate in the
energy range 3-30 keV is, after background subtraction, $\sim$ 37 c
s$^{-1}$.  A FFT was applied to each of two series of 8192 data points
accumulated with a time resolution of 0.366 s, and the result averaged
to produce the power density spectrum (PDS) of Fig \ref{pds992}.  A
modulation at 0.031 Hz and several of its harmonics are clearly seen.
In order to measure the pulsation period, an epoch
folding search was performed near to the period expected from the FFT power
spectrum.  A solar barycentric pulse period of 31.9 s was obtained
from the resulting pulse-folding periodogram.  The pulse profile
obtained from this period was used as a template.  Then, both the
observations and the template were divided into 5 roughly equal time
intervals of 676 s each ($\sim$ 21 cycles).  The difference between
the actual pulse period at the epoch of observation and the period
used to fold the data was determined by cross-correlating the original
light curve and the template.  The $P_{actual}-P_{template}$ shifts
were fitted to a linear function to obtain an improved period.  A new
template was derived for this period and the process repeated.  The
derived pulse period was 31.8851$\pm$0.0004 s.  The error represents
the scatter of the points about the best-fit straight line.

\begin{figure}
    \begin{center}
    \leavevmode
\epsfig{file=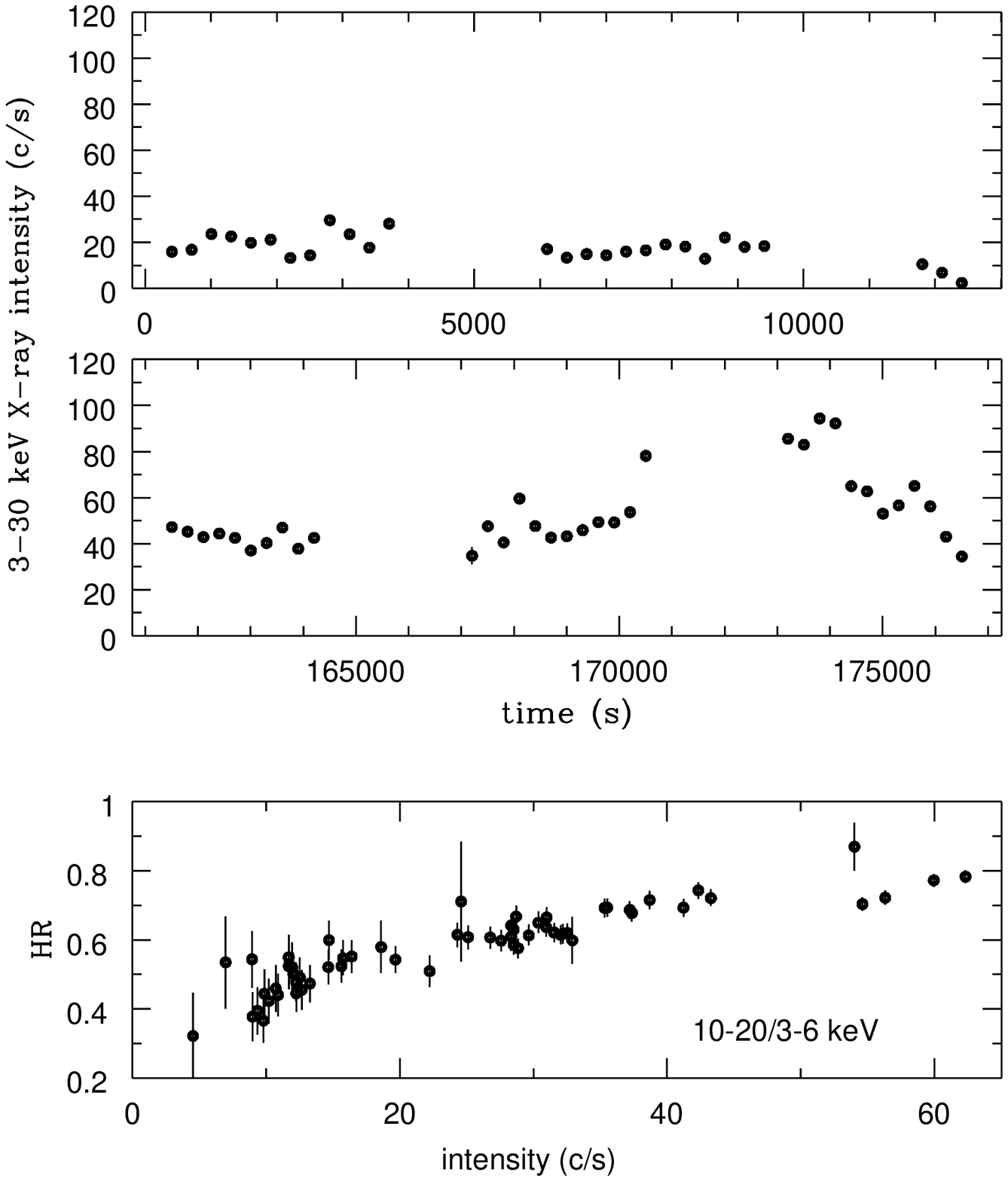, width=8cm, bbllx=75pt, bblly=180pt,
  bburx=545pt, bbury=715pt, clip=}
 \end{center}
        \caption{{\em Upper panels: }Lightcurve of LS 992/RX J0812.4--3114 
covering the entire observation.  Significant variability is seen on
timescales of hours. Time 0 is JD 2,450,845.546 and the bin size is
300 s. The scale of the Y-axis in the upper panel has been kept the
same as the lower panel to allow direct comparison. {\em Lower panel}
Hardness ratio as a function of intensity.}
        \label{lc_whole992}
        \end{figure}

LS 992/RX J0812.4--3114 shows significant variability on timescales of
a few hours, with variations in the amplitude of almost an order of
magnitude (see Fig \ref{lc_whole992}) which cannot be attributed to
the pulsations.  Motch et al.  (1997) reported a decline in X-ray flux
for this source of a factor $\sim$100 between the ROSAT survey observations
and the follow-up pointed observations over a year later.  This
amplitude of variation is typical of BeXRBs.

        \begin{figure}
    \begin{center}
    \leavevmode
\epsfig{file=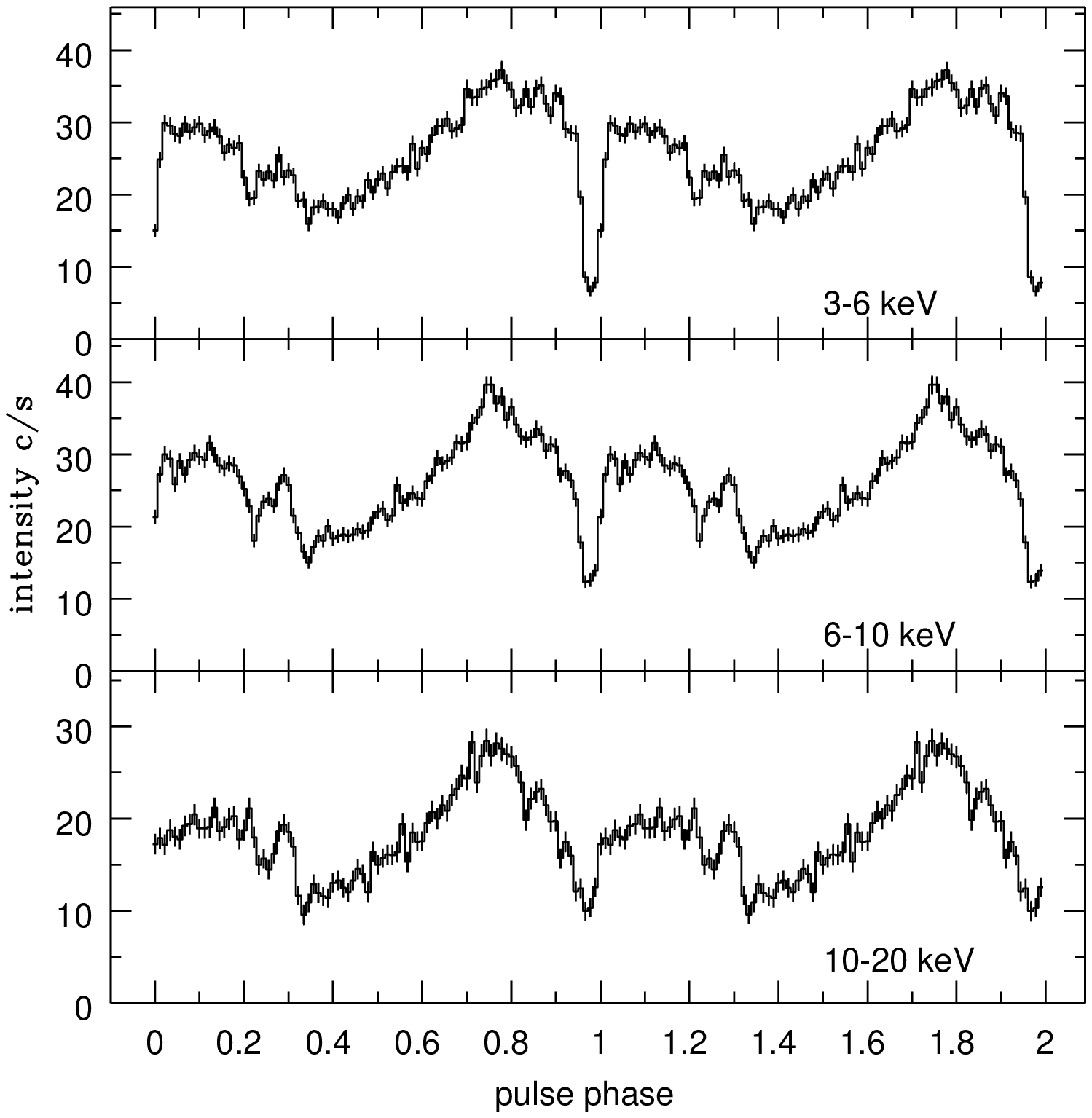, width=8cm, bbllx=55pt, bblly=260pt,
  bburx=504pt, bbury=711pt, clip=}
 \end{center}
        \caption{Pulse profiles of LS 992/RX J0812.4--3114 at 3-6, 6-10
and 10-20 keV. Note the sharp absorption feature at relative pulse
phase $\sim$0.95}
        \label{pp992}
        \end{figure}

Fig \ref{lc_whole992} also shows the hardness ratio in the
photon-energy bands 10-20 and 3-6 keV as a function of the summed
count rate of the two bands.  A positive correlation is observed, the
spectrum becoming harder with increasing intensity, the hardness
ratio increasing from 0.3 at $\sim$ 5 c s$^{-1}$ to 0.8 at $\sim$ 60 c
s$^{-1}$, behaviour typical of X-ray pulsar systems in general.

The pulse profiles folded with the best-fit period are shown in Fig
\ref{pp992} for the 3-6 keV, 6-10 keV and 10-20 keV energy bands.  The
three pulse profiles clearly do not resemble sinusoidal curves, as
expected from the presence of the harmonics in the PDS.  They consist
of a broad and irregular peak divided into two halves by a sharp
absorption feature.  The strength of this dip decreases with
increasing energy.  The peak appears to be flatter at low energies.

The pulse fraction, which we have defined here as 
PF = ($I_{max}$--$I_{min}$)/($I_{max}$+$I_{min}$), decreases as we progress
into higher energies: $\sim$ 70\%, 51\% and 47\% for the low, middle
and high energy intervals.  $I_{max}$ and $I_{min}$ are the counts per
second at pulse peak and pulse dip.

\section{Spectral Analysis}
\label{spectral}

\begin{figure}
    \begin{center}
    \leavevmode
    \epsfig{file=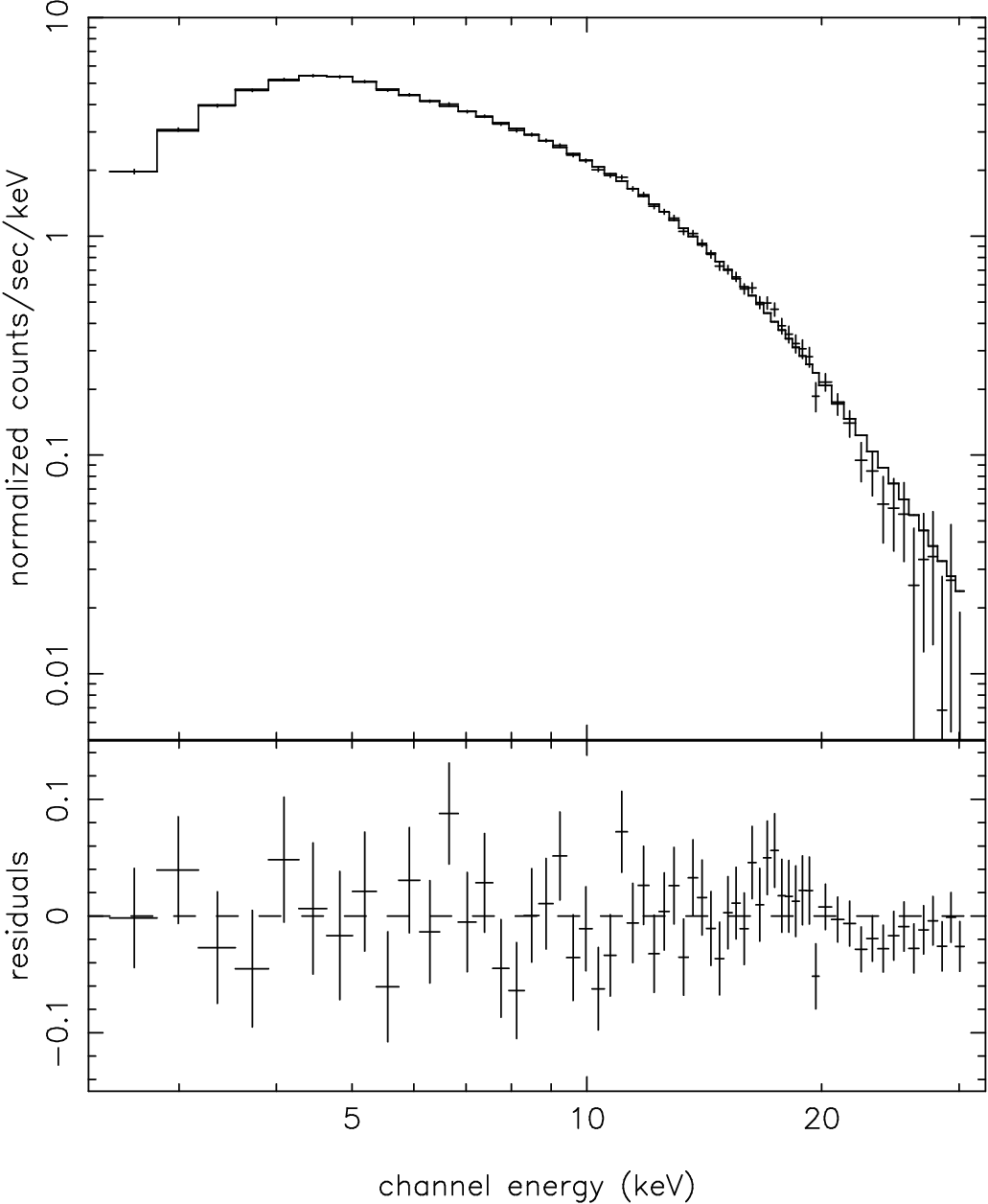, width=7cm, bbllx=34pt, bblly=55pt,
  bburx=540pt, bbury=665pt, clip=}
 \end{center}
        \caption{PCA spectrum of LS 992/RX J0812.4-3114. The straight line 
represents the best-model fit, a power-law plus exponential cut-off. The 
residuals are shown in the lower panel}
        \label{sp992}
        \end{figure}

In order to investigate the X-ray energy emission of LS 992/RX
J0812.4--3114, fits using a variety of models were performed.  Table 1
gives the results of the spectral analysis when the entire observation
is considered.  The power-law plus high-energy cut-off model gave the
lowest reduced $\chi^2$ ($\chi^2_r$=1.00 for 56 dof) for a cut-off
energy of 4.9$\pm$0.4 keV.  The best-fit value of the photon index
$\alpha$ is 1.0$\pm$0.1 and low-energy absorption, N$_H$, equivalent
to $\sim$0.5$\pm$0.3 $\times$ 10$^{22}$ cm$^{-2}$ is required (see Fig
\ref{sp992}).  This value is consistent (Ryter, Ceasarsky \& Audouze
1975) with a colour excess of E(B--V)=0.70 (Motch et al.  1997) and
indicates that the absorption is entirely due to the insterstellar
medium and that very little circumstellar matter exists.  This fact
would then explain the lack of an iron line.

Since the HR varied significantly throughout the observation,
indicating some spectral changes with X-ray intensity, we divided the
observation up into two different intervals and a spectrum for each
interval was obtained. The difference in the mean 3-30 keV luminosity
between the intervals is a factor $\sim$ 3.5. Whilst the value of
$\alpha$ remained constant at 1.0, the cut-off and absorption showed a
slightly higher value at higher count rate, varying from 4.5$\pm$0.4
keV and 0.2 $\times$ 10$^{22}$ cm$^{-2}$ at L$_x$ $\sim$4.7 $\times$
10$^{35}$ erg s$^{-1}$, to 5.5$\pm$0.5 keV and 0.8 $\times$ 10$^{22}$
cm$^{-2}$ at L$_x$ $\sim$1.6 $\times$ 10$^{36}$ erg s$^{-1}$. The
higher N$_{H}$ at higher luminosities supports the idea that the flare
is due to enhanced accretion from the stellar wind/disc of the Be
star.

The observed cut-off energy is lower than the typical value found in
other X-ray pulsars.  However, Reynolds, Parmar \& White (1993) showed
a correlation between the cut-off energy and the X-ray luminosity in
the BeXRB EXO 2030+375, with E$_{cut}$ $\sim$4.7 keV for a L$_x$
$\sim$1.2 $\times$ 10$^{36}$ erg s$^{-1}$.  Therefore the value of 4.9
keV may be a consequence of the low luminosity observed in LS 992/RX
J0812.4--3114.  The X-ray luminosity is $\sim$1.1 $\times$ 10$^{36}$
erg s$^{-1}$ in the energy range 3-30 keV assuming a distance of $\sim$9
kpc.  The two blackbody component model may be disregarded as
unrealistic, since it gives a hydrogen column density of 0.0
cm$^{-2}$, whereas the power-law plus blackbody component model gives
a radius for the emitting area much lower than the minimum accepted
radius of the polar caps of $\sim$1 km.

\begin{table}
\begin{center}
\label{models}
\caption{Spectral fits results. Uncertainties are 90\% confidence. The
spectrum was fitted in the energy range 2.7-30 keV.}
\begin{tabular}{lc}
\hline
paramters				&  value\\
\hline
\multicolumn{2}{l}{{\bf Power-law \& blackbody}}\\ 
N$_H$ (10$^{22}$ atoms cm$^{-2}$) 	& 3.4$\pm$0.5		\\
$\alpha$				& 1.96$\pm$0.08		\\
kT (keV)				& 2.7$\pm$0.2		\\
R (km)					& 0.21$\pm$0.02		\\
$\chi^2_r$(dof) 			& 1.59(56)		\\
\hline
\multicolumn{2}{l}{{\bf Two blackbody}}\\
kT$_1$ (keV)				& 1.27$\pm$0.03		\\
R$_1$ (km)				& 1.08$\pm$0.03		\\
kT$_2$ (keV)				& 3.57$\pm$0.09		\\
R$_2$ (km)				& 0.20$\pm$0.01		\\
N$_H$ (10$^{22}$ atoms cm$^{-2}$) 	& -			\\
$\chi^2_r$(dof) 			& 1.53(57)		\\
\hline
\multicolumn{2}{l}{{\bf Cut-off power-law }}\\
N$_H$ (10$^{22}$ atoms cm$^{-2}$)	& 0.5$\pm$0.3		\\
$\alpha$				& 1.0$\pm$0.1		\\
E$_{cut}$				& 4.9$\pm$0.4		\\
E$_{fold}$ (keV)			& 11.6$\pm$1.3		\\
$\chi^2_r$(dof) 			& 1.00(56)		\\
\hline
\end{tabular}
\end{center}
\end{table}

\section{Pulse phase-resolved spectroscopy}

In order to investigate the pulse phase dependence, 5 phase-resolved
energy spectra were produced, and each fitted with the power-law plus
cut-off energy model used in Sect.  \ref{spectral}.  To improve the
signal to noise ratio we used data only from the top layer anode from
the detectors.  We also restricted the energy interval to 2.7-20 keV.

\begin{figure}
    \begin{center}
    \leavevmode
    \epsfig{file=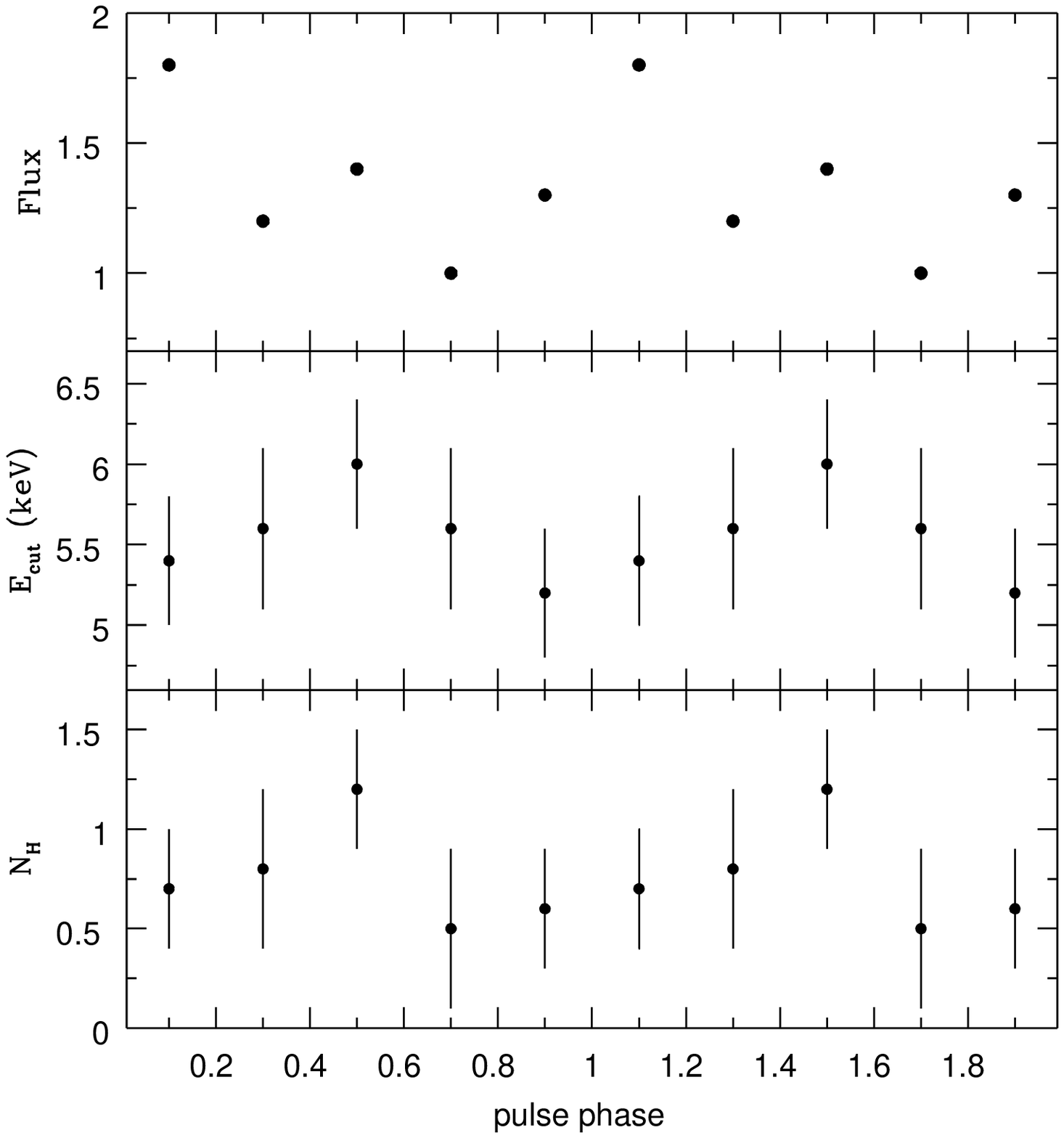, width=7cm, bbllx=45pt, bblly=230pt,
  bburx=505pt, bbury=710pt, clip=}
 \end{center}
        \caption{X-ray spectral parameters as a function of the pulse phase. 
Errors are given at 90\% confidence interval. The flux in the top panel is in 
units of 10$^{-10}$ erg cm$^{-2}$ s$^{-1}$ and corresponds to the energy range 
3-20 keV. The values of N$_H$ are given in units of 10$^{22}$ cm$^{-2}$. Two 
pulse phases are shown for clarity}
        \label{pps}
        \end{figure}

Initially, in order to secure enough counts we considered the entire
observation. However, it was not possible to simultaneously constrain
the power-law index $\alpha$, the cut-off energy E$_{cut}$ and the
absorption column N$_H$. This inconsistency in the data may be
attributed to the variation of the X-ray intensity observed throughout
the observation. That is, the spectral parameters during the ``flare''
event (middle panel of Fig \ref{lc_whole992}) show different values to
those obtained during the ``quiescent'' state (upper panel of Fig
\ref{lc_whole992}), except the photon index which remained fairly
constant. Consequently, the results presented in Table 2 and Fig
\ref{pps} correspond to the second interval only (``flare'' event).

In spite of the complex pulse profile, no correlation could be found
between the X-ray flux in the energy range 3-20 keV and the absorption
column and cut-off energy.  Nevertheless, these two parameters do vary
with pulse phase, as can be seen in Fig \ref{pps}.

\begin{table}
\begin{center}
\label{models}
\caption{Spectral parameters as a function of the pulse phase. Uncertainties are 
90\% confidence. The photon index was fixed to $\alpha$=1.04}
\begin{tabular}{lccccc}
\hline
phase	&Flux$^a$  &N$_H^b$	&E$_{cut}$	&E$_{fold}$	&$\chi^2$(dof)\\
     &(3-20 keV)   &		& (keV)		& (keV)		&\\
\hline
0.1  	&1.8  	&0.7$\pm$0.3  	&5.4$\pm$0.4  	&14.9$\pm$0.8  	&1.16(44) \\
0.3  	&1.2  	&0.8$\pm$0.4  	&5.6$\pm$0.5  	&11.2$\pm$0.6  	&1.19(44) \\
0.5  	&1.4  	&1.2$\pm$0.3  	&6.0$\pm$0.4  	&13.0$\pm$1.0  	&1.08(44) \\
0.7  	&1.0  	&0.5$\pm$0.4  	&5.6$\pm$0.5  	&13.3$\pm$1.0  	&0.97(44) \\
0.9  	&1.3  	&0.6$\pm$0.3  	&5.2$\pm$0.4  	&15.0$\pm$1.0  	&1.02(44) \\
\hline
\end{tabular}
\end{center}
$a$: in units of 10$^{-10}$ erg s$^{-1}$ cm$^{-2}$ \\
$b$: in units of 10$^{22}$ cm$^{-2}$\\
\end{table}

\section{Discussion}

Pulsations have previously been found in both high and low mass X-ray
binaries.  However, of the $\sim$65 X-ray pulsars currently known, only 5
have been identified as LMXRBs, whereas $\sim$60 have an OB-type star as the
optical counterpart.  For the confirmed BeXRB systems, around 70$\%$
are found to be pulsators, rising to 75-80\% if suspected BeXRBs are
included. The periods of X-ray pulsars are distributed over a factor
$\sim$10$^5$, from 2.5 ms (SAX J1808--369, the bursting millisecond
X-ray pulsar) to 1412 s (LS\,I +61 235, a persistent BeXRB), with no
evidence for clustering at any particular period. The range for BeXRBs
is almost as extensive, the fastest being 69 ms (A 0535--668). 

LS 992/RX J0812.4--3114 appears to be a typical transient BeXRB.  This
source was active during the ROSAT survey observations reported by
Motch et al. (1997), but those authors failed to detect the source
during a pointed ROSAT PSPC observation in 1992 November 20,
indicating that the X-ray emission decreased by about a factor of
$\sim$100.  During the RXTE observation, however, we obtained an
extrapolated X-ray luminosity in the energy range covered by ROSAT
(0.1-2.4 keV) of 9.1 $\times$ 10$^{34}$ erg s$^{-1}$, similar to the
maximum reported by Motch et al.  (1997), 1.3 $\times$ 10$^{35}$ erg
s$^{-1}$.  In the energy range 3-30 keV LS 992/RXJ0812.4--3114
displayed an X-ray luminosity of 1.1 $\times$ 10$^{36}$ erg s$^{-1}$
for an assumed distance of $\sim$9 kpc.

In addition to this long-term variability, LS 992/RX J0812.4--3114
also shows changes in the X-ray intensity on short time scales.  Fig
\ref{lc_whole992} displays the lightcurve of the entire observation,
where X-ray intensity variations of one order of magnitude are seen on
a time scale of a few hours.  LS 992/RX J0812.4--3114 was found to be
one of the brightest and hardest new BeXRB candidates in the ROSAT
all-sky survey (0.1-2.4 keV) (Motch et al.  1997).

The sharp minimum that appears in the pulse profiles may be
interpreted as absorption by the accretion flow.  Cemelji\'c \& Bulik
(1998) have shown that X-rays produced near the surface of the
neutron star may be eclipsed at a certain phase due to the passage of
the accretion flow through the line of sight as the compact star
rotates. The greater penetration of harder X-rays would explain the
decrease in the strength of the peak with energy.

At first glance, the low absorption column N$_H$=0.5 $\times$
10$^{22}$ cm$^{-2}$ and the classification of the system as a
relatively bright V=12.4 Be star may seem not to agree with a distance
of 9 kpc.  However, we note that the location of LS 992/RX
J0812.4--3114 (l=249.6, b=1.55) within Puppis shows a remarkable lack
of interstellar reddening (Lucke 1978).  Wilson \& FitzGerald (1972)
studied the distribution of OB stars and color excesses in the region
of Puppis (l=245) and find very low excesses even out to 5 or 6
kpc. The observed E(B--V)=0.70 and m$_v$=12.4 suggests an absolute
magnitude M$_v$$\sim$--4.6 for a distance of 9 kpc, consistent with
an O9.5V star (e.g. Deutschman, Davis \& Schild 1976, Wegner 1994).
Therefore, a distance of 9 kpc is not unreasonable for the assumed Be
star identification.

\section{Conclusion}

We have carried out X-ray timing and spectral analyses of the newly
discovered high mass X-ray binary LS 992/RX J0812.4--3114.  The
timing analysis has resulted in the detection of pulsations with a
barycentric spin period of 31.8851$\pm$0.0004 s.  Like many other
X-ray pulsars the energy spectrum is well represented by a power-law
component, modified at energies above a high energy cut-off E$_{cut}$
by an exponential function with folding energy E$_{fold}$ and at low
energies by photoelectric absorption due to intervening cold matter of
density N$_H$ $\sim$ 0.5 $\times$ 10$^{22}$ cm$^{-2}$.  No iron line at
$\sim$6.4 keV is required.  Thus LS 992/RX J0812.4--3114 appears
to be a typical transient BeXRB, with an expected orbital period in
the region of tens of days according to the Corbet diagram (Corbet
1986).

\subsection*{Acknowledgments}

P. Reig acknowledges partial support via the European Union Training and 
Mobility of Researchers Network Grant ERBFMRX/CP98/0195.

\bsp

\end{document}